%% file: triangularJ1J2Jchi_AML_3.tex
\documentclass[aps,prl,reprint,longbibliography]{revtex4-1}
\usepackage{times}
\usepackage[colorlinks=true, urlcolor=blue, linkcolor=red, citecolor=blue, pdftex]{hyperref}
\usepackage{graphicx}
\usepackage{amsmath}
\usepackage{amssymb}
\usepackage{color}

\def\ket#1{\left|#1 \right\rangle}

\def\braket#1#2{\left\langle #1 | #2 \right\rangle}
\def\matrix22#1#2#3#4{\left(\begin{array}{cc}#1&#2\\#3&#4\end{array}\right)}

\begin{document}
\title{{Chiral Spin Liquid and Quantum Criticality in Extended $S=1/2$ Heisenberg Models on the Triangular Lattice.}}
\author{Alexander Wietek}
\email{alexander.wietek@uibk.ac.at}
\author{Andreas M. L\"auchli}
\affiliation{Institut f\"ur Theoretische Physik, Universit\"at Innsbruck, A-6020 Innsbruck, Austria}
\date{\today}
\begin{abstract}
{ We investigate the $J_1$-$J_2$ Heisenberg model on the triangular
lattice with an additional scalar chirality term and show that a chiral
spin liquid is stabilized in a sizeable region of the phase
diagram. This topological phase is situated in between a
coplanar $120^\circ$ N\'{e}el ordered and a non-coplanar tetrahedrally
ordered phase. Furthermore we discuss the nature of the spin-disordered 
intermediate phase in the $J_1$-$J_2$ model. We compare the groundstates from
Exact Diagonalization with a Dirac spin liquid wavefunction
and propose a scenario where this wavefunction describes the quantum
critical point between the $120^\circ$ magnetically ordered phase 
and a putative $\mathbb{Z}_2$ spin liquid.}
\end{abstract}

\maketitle
\paragraph{Introduction ---}
The emergence of quantum spin liquids in frustrated quantum 
magnetism is an exciting phenomenon in contemporary condensed matter
physics \cite{Balents2010}. These novel states of matter
exhibit fascinating properties such as long-range groundstate
entanglement \cite{Kitaev2006,Levin2006} or
anyonic braiding statistics of quasiparticle excitations, relevant for
a potential implementation of topological quantum computation
\cite{Nayak2008}. Only very recently such phases have been found
to be stabilized in realistic local spin 
models~\cite{Gong2014,Bauer2014,Wietek2015,Hickey2015,He2013,Nataf2016,Kumar2015,
Gorohovsky2015, Thomale2009,Nielsen2012,Meng2015,Sachdev1992,Moessner2001,Misguich2002,Kitaev20062}.

Triangular lattice Heisenberg models are a paradigm of frustrated
magnetism. Although the Heisenberg model with only nearest
neighbour interaction is known to stabilize a regular $120^\circ$
N\'{e}el order \cite{Jolicoeur1990,Bernu1992,Capriotti1999,White2007} adding further 
interaction terms may increase frustration and induce magnetic 
disorder to the system. 
Experimentally, several materials with triangular lattice geometry
do not exhibit any sign of magnetic ordering down to 
lowest temperatures \cite{Kurosaki2005, Shimizu2003,
  Yamashita2010,Itou2008}. These include 
for example the organic Mott insulators like
$\mathrm{\kappa-(BEDT-TTF)_2Cu_2(CN)_3}$ \cite{Kurosaki2005, Shimizu2003}
or $\mathrm{EtMe_3Sb[Pd(dmit)_2]_2}$ \cite{Yamashita2010,Itou2008}
and are thus candidates realizing spin liquid physics. 

Historically Kalmeyer and Laughlin \cite{Kalmeyer1987} introduced 
the \textit{chiral spin liquid} (CSL) state on the triangular lattice. This 
state closely related to the celebrated Laughlin wavefunction of
the fractional quantum Hall effect has recently been shown to 
be the ground state of several extended Heisenberg models
on the kagom\'{e} lattice \cite{He2013,Gong2014,Bauer2014,Wietek2015}. 
The question arises whether a CSL
can indeed be realized on the triangular lattice as originally
proposed. In a recent study \cite{Nataf2016}
this was shown for SU($N$) models for $N\geq 3$.
In this letter we provide conclusive evidence that indeed the CSL is
stabilized in a spin-$1/2$ Heisenberg model upon adding a further
scalar chirality term $J_\chi\vec{S}_i\cdot(\vec{S}_j\times \vec{S}_k)$ 
similar as in Refs.~\cite{Bauer2014,Wietek2015,Hickey2015,Nataf2016}.
Such a term can be realized as a lowest order effective Heisenberg Hamiltonian
of the Hubbard model upon adding $\Phi$ flux through the elementary
plaquettes \cite{Motrunich2006, Bauer2014},
either via a magnetic field or by introducing artificial
gauge fields in possible cold atoms experiments 
\cite{Aidelsburger2013, Miyake2013}. The coupling 
constants then relate to the Hubbard model
parameters $t$ and $U$ as $J_1 \sim t^2/U$ and $J_\chi \sim \Phi
t^3/U^2$ where $J_1$ (resp. $J_\chi$) is the nearest neighbour
Heisenberg (resp. scalar chirality) coupling.

Another open question in frustrated magnetism of the triangular lattice 
is the nature of the intermediate phase in the phase diagram of the $S=1/2$
Heisenberg model with added next-nearest neighbour couplings around $J_2
/ J_1 \approx 1/8$.
Several authors \cite{Jolicoeur1990,Chubukov1992,Lecheminant1995} found
a spin disordered state. Recently several numerical studies~\cite{Kaneko2014,Zhu2015,Saadatmand2015,Hu2015a,Iqbal2016}
proposed that a topological spin liquid state of some kind might be realized in this regime.
The exact nature of this phase yet remains unclear. In this Letter
we advocate the presence of a $O(4)^*$ quantum critical point~\cite{Chubukov1994a,Chubukov1994b,Whitsitt2016} 
separating the $120^\circ$ N\'eel order from a putative $\mathbb{Z}_2$ spin liquid. The diverging 
correlation length at this quantum critical point and the neighbouring first order phase 
transition into the stripy collinear magnetic ordered phase render the unambiguous identification of 
the intermediate spin liquid phase challenging however. 

\begin{figure}[b]
  \centering
  \includegraphics[width=\linewidth]{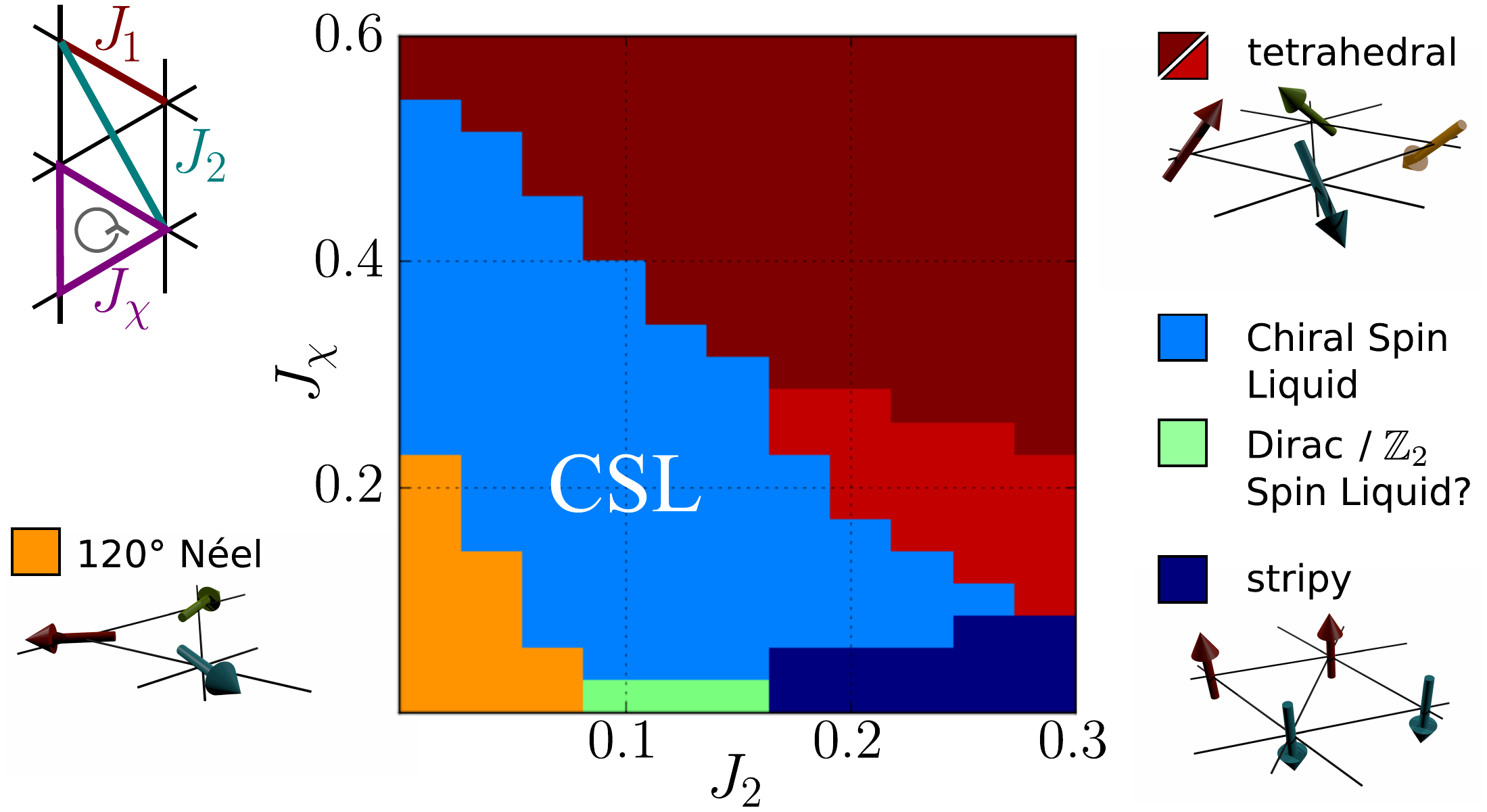}
  \caption{Approximate $T=0$ phase diagram of the $J_1$-$J_2$-$J_\chi$ model
   on the triangular lattice, c.f.~Eq.~\eqref{eq:hamiltonian}. The extent of phases is 
   inferred from excitation spectra from ED on a periodic $36$ sites 
   triangular simulation cluster, see main text for details.
  } 
  \label{fig:phasediagram}
\end{figure}
\paragraph{Model ---}
We investigate the Heisenberg model with
nearest and next-nearest neighbour interactions with an additional 
uniform scalar chirality term on the triangular lattice
\begin{align}
  \begin{split}
    \label{eq:hamiltonian}
    \mathcal{H} \,=\, &J_1\sum\limits_{\left< i,j\right>}\vec{S}_i\cdot\vec{S}_j \ + 
    J_2\sum\limits_{\left<\left<  i,j\right>\right>}\vec{S}_i\cdot\vec{S}_j + \\
    &J_{\chi}\sum\limits_{i,j,k \in \bigtriangleup} \vec{S}_i\cdot(\vec{S}_j\times \vec{S}_k)
  \end{split}
\end{align}
where we set $J_1 \equiv 1$ and consider $J_2,J_\chi \ge 0$. Amongst
a $120^\circ$ N\'{e}el order, a stripy and a tetrahedral magnetic
order we find a CSL being realized in an extended region of the phase 
diagram in Fig.~\ref{fig:phasediagram}. 
A first study of the classical phase diagram for $J_\chi = 0$
\cite{Jolicoeur1990} found a three sublattice $120^\circ$
N\'{e}el ordered groundstate for $J_2 < 1/8$ whereas
for  $1/8<J_2 < 1$ a two-parameter family of magnetic ground states with a
four-site unit cell was found~\cite{Chubukov1992}. Two high-symmetry solutions 
within this manifold are a two-sublattice collinear stripy magnetic order breaking 
lattice rotation symmetry and a tetrahedral non-coplanar state with a uniform scalar 
spin chirality on all triangles.  Taking into account quantum fluctuations
by applying spin-wave theory, large-$S$ perturbation theory and ED studies 
\cite{Jolicoeur1990,Chubukov1992,Lecheminant1995} the 
degeneracy is lifted by an \textit{order-by-disorder} mechanism. The true quantum groundstate 
for $ J_2 \gtrsim 0.18 $ exhibits stripy N\'{e}el order. 
Yet the behaviour of the system close to the classical phase transition point $J_2 = 1/8$ has not been fully understood. 

\paragraph{Phase diagram ---}
We performed ED calculations on a $N_s = 36$
sites simulation cluster with periodic boundary conditions to
investigate ground state properties and order parameters of the model
\eqref{eq:hamiltonian}. We have also checked selected results on smaller
clusters, but the $N_s=36$ cluster is particularly well suited because this 
single cluster can harbour all phases which we were able to detect.

\begin{figure}[t!]
  \centerline{\includegraphics[width=\linewidth]{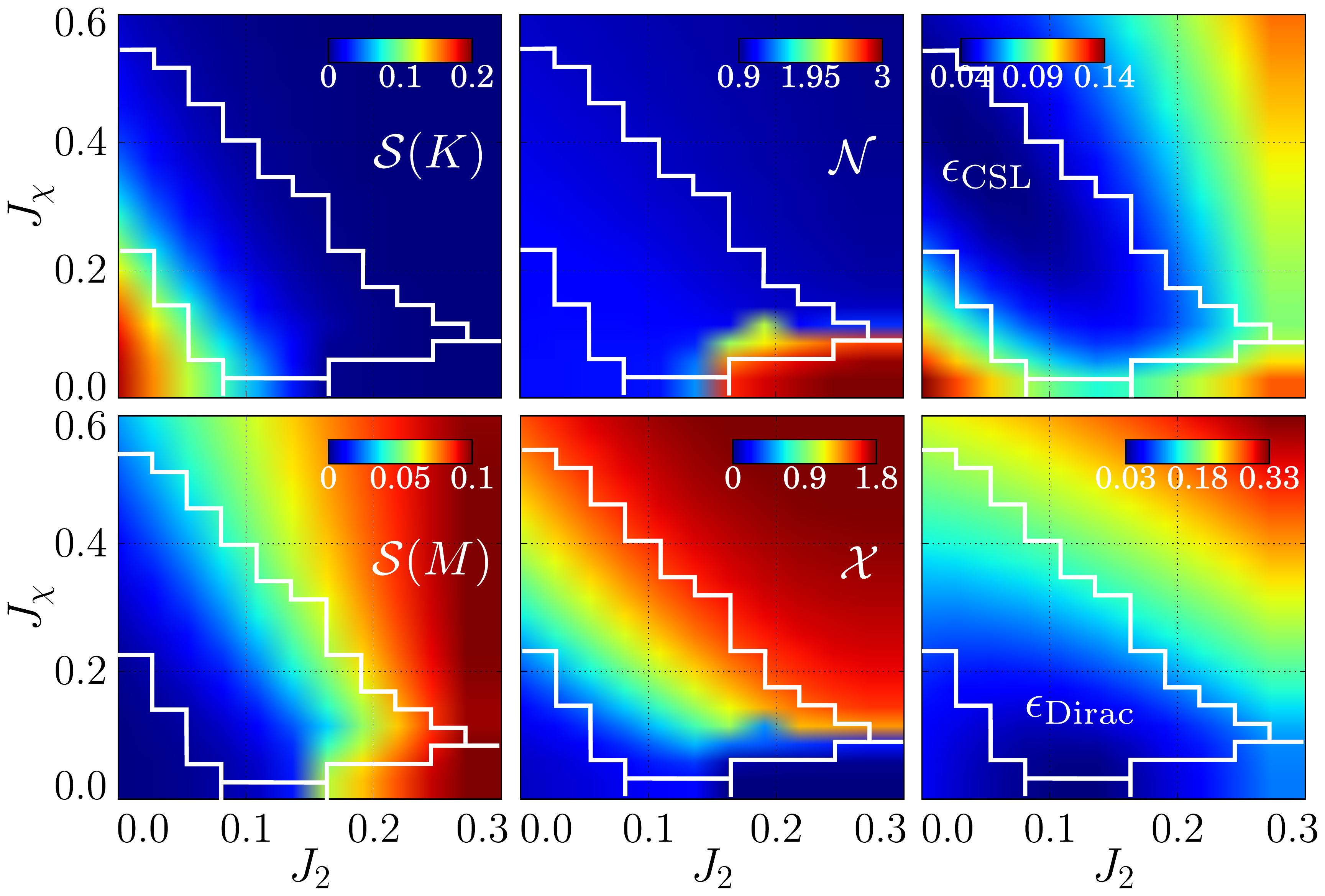}}
  \caption{Order parameters and variational energies of model
    wavefunctions. 
    \textit{Left:} static spin structure factor
    $\mathcal{S}(q)$. 
    evaluated at $K$ and $M$ point. 
    \textit{Middle:} nematic order parameter $\mathcal{N}$ as in
    Eq.~\eqref{eq:nematicorderparameter} and (disconnected) scalar chirality
    correlation $\mathcal{X}$ as in
    Eq.~\eqref{eq:chiralityorderparameter}
    \textit{Right:} Variational energies $\epsilon =
    (E_\textnormal{model} -E_\textnormal{ED} )/E_\textnormal{ED} $ for
    the chiral and Dirac spin liquid. }
  \label{fig:orderparameters}
\end{figure}

We present the approximate phase diagram in Fig.~\ref{fig:phasediagram} based on
the quantum numbers of the ground state level and the first excited state. The groundstate 
is always in the $\Gamma$.A1 representation (except in the stripy phase
where $\Gamma$.A1 and the two $\Gamma$.E2
sectors are almost degenerate). The symmetry sector of the first
excited state determines the phase.
\textit{Orange}: $S=1$ $K$.A1 ($120^\circ$ N\'{e}el)
\textit{Light blue}: $S=0$ $\Gamma$.E2b (CSL), 
\textit{Green}: $S=0$ $\Gamma$.E2a,$\Gamma$.E2b degenerate (Dirac/$\mathbb{Z}_2$ spin liquid), 
\textit{Dark Blue}: $S=0$ $\Gamma$.A1, $\Gamma$.E2a, $\Gamma$.E2b 
degenerate (stripy magnetic order),
\textit{Dark red/Light red}: $S=1$ $M$.A / $S=0$ $\Gamma$.E2a (tetrahedral magnetic order)
For the magnetically ordered phases these quantum numbers follow from a standard tower of states symmetry
analysis~\cite{Lhuillier2005,Rousochatzakis2008}. 
The spectral phase diagram is further corroborated by the analysis of relevant order parameters and variational
energies of model wave functions, c.f.~Fig.~\ref{fig:orderparameters}, where the agreement is striking.
We now proceed to a detailed discussion of the phases and the corresponding order parameters.

\textit{$120^\circ$ N\'{e}el order:} At the Heisenberg point $J_2 =
J_\chi = 0$ the system exhibits $120^\circ$ N\'{e}el
order \cite{Lecheminant1995} for which the static spin structure factor 
 $ \mathcal{S}(q) = |\sum_{j}
 e^{iq(\underline{r}_j-\underline{r}_0)}\langle \vec{S}_j\cdot\vec{S}_0
\rangle |^2/N_s$
is peaked at the $K$-point in the Brillouin zone \cite{Messio2011}. 
The Anderson tower of states for this ordering \cite{Bernu1992}
yields spin-$1$ excitations with symmetry sectors $K$.A1 and
$\Gamma$.B2. In the orange region in Fig.\ref{fig:phasediagram}
the first excited state is a triplet and belongs to the $K$.A1
representation. Here also the structure factor of the groundstate
evaluated at the $K$-point is peaked, cf. Fig.~\ref{fig:orderparameters}. Thus this region
determines the approximate extent of the $120^\circ$ N\'{e}el phase. 

\textit{Stripy order} is characterized by spins aligned
ferromagnetically along one direction of the triangular lattice and
antiferromagnetically along the other two (c.f. illustration in Fig.~\ref{fig:phasediagram}). It breaks 
SU($2$) spin rotation symmetry and discrete lattice rotation symmetry.
According to \cite{Lecheminant1995,Jolicoeur1990,Chubukov1992}
the stripy order is stabilized for $J_\chi=0$ and $J_2 \gtrsim
0.18$. The groundstate in the singlet sector is 
expected to be threefold degenerate with corresponding irreducible
representations in the $\Gamma$.A1 and the two dimensional 
$\Gamma$.E2 singlet sectors \cite{Lecheminant1995}.
The area where those three states are nearly degenerate is coloured
dark blue in Fig.~\ref{fig:phasediagram}.
The spin structure factor  $ \mathcal{S}(q)$ 
is  peaked at the $M$-point for both the stripy and the tetrahedral phase 
\cite{Messio2011}. 
We computed the nematic order parameter
\begin{equation}
  \label{eq:nematicorderparameter}
  \mathcal{N} = \sum\limits_{(i,j) \parallel (0,1)} 
  \left< \left(\vec{S}_0\cdot\vec{S}_1 \right)\left(\vec{S}_i\cdot\vec{S}_j \right)\right> _c
\end{equation}
as a sum over nearest neighbour connected dimer-dimer correlations where
only parallel and non-overlapping dimer configurations are considered, Fig.~\ref{fig:orderparameters}. 
The region where this order parameter is large
coincides with the dark blue region in Fig.~\ref{fig:phasediagram}, 
hence, the approximate extent of the stripy phase.

\textit{Tetrahedral order} is a non-coplanar regular magnetic 
order~\cite{Kubo1997,Messio2011,Hickey2015}. 
The spins in a four-site unit cell are arranged in a way that their orientations form
a tetrahedron and span a finite volume on each triangle. The expectation value
of the scalar chirality is thus non-zero and uniform for the classical spin configuration.
This implies significant summed scalar chirality correlations 
\begin{equation}
  \label{eq:chiralityorderparameter}
  \mathcal{X} = \sum\limits_{(i,j,k) \in \bigtriangleup}\left< \chi_{(0,1,2)}\cdot\chi_{(i,j,k)}\right> 
\end{equation}
where $\chi_{(i,j,k)} = \vec{S}_i\cdot(\vec{S}_j\times \vec{S}_k)$ and the sum runs over all 
non-overlapping triangles. 
Tetrahedral order is classically degenerate with the stripy order when $J_\chi=0$ and
$1/8 < J_2<1$ \cite{Jolicoeur1990,Chubukov1992,Lecheminant1995}. 
Therefore we expect that this state will be energetically favored
over the stripy phase upon adding a scalar chirality term in the
Hamiltonian. 
The tetrahedral state does not break lattice
rotation symmetry and the nematic order parameter
\eqref{eq:nematicorderparameter} is relatively small
 in Fig.~\ref{fig:orderparameters}. We find a sharp
transition between the nematic order parameter and scalar chirality correlations. 
Moreover a level crossing at a finite angle in the 
excitation spectra strongly indicates that this is a first order phase transition.
Tower of states analysis predicts that 
the $S=1$ levels belong to the irreducible representation $M$.A.
In most of the region where both the 
structure factor at $M$ and the summed scalar chirality correlations 
$\mathcal{X}$ are large we find that the first excited level above the 
groundstate belongs to this representation. This region is coloured dark red in
Fig~\ref{fig:phasediagram}.  Close to the stripy phase we observe 
that the first excited level is a $S=0$ $\Gamma$.E2a level, shown as the 
light red region in Fig.~\ref{fig:phasediagram}. We believe
that this level is an artifact of the finite size sample and is
related to the order by disorder mechanism. In neither of the groundstate 
correlation functions we can see a difference between the light red region 
and the red region and thus conclude that also this 
region belongs to the same tetrahedral phase.

\paragraph{Chiral Spin Liquids ---}
are spin disordered chiral topological states. 
Hallmark features of this phase are the topology dependent ground state
degeneracy, long-range entanglement, abelian anyonic excitations and 
gapless chiral edge modes. Several instances of this phase have recently 
been found in local spin models~\cite{He2013,Gong2014,Bauer2014,Wietek2015,Hickey2015,Nataf2016,Kumar2015,Gorohovsky2015, Thomale2009,Nielsen2012,Meng2015}.
It has been understood that a representative lattice model wave function for the CSL
is provided by Gutzwiller projected parton wavefunctions (GPWF) with a completely 
filled parton band with Chern number~$\pm1$~\cite{Wen2002,Wen2004,Wietek2015,Nataf2016}.
We observe no strong magnetic structure peak in between 
the $120^\circ$ N\'{e}el order and the tetrahedral, cf. Fig.~\ref{fig:orderparameters}.
 Therefore a spin disordered state is formed in a sizeable
 intermediate region. The summed scalar chirality correlations $\mathcal{X}$
in Fig.~\ref{fig:orderparameters} are relatively large in this regime compliant with the fact that here a CSL
with a uniform chirality is formed. We will now show conclusive evidence that this is indeed the case. 
We do so by constructing two GPWFs describing the two topological sectors of the
chiral spin liquid on the torus and by computing their overlaps with the two
lowest lying exact eigenstates from ED, similarly as in refs.~\cite{Nataf2016,Wietek2015}. 
\begin{figure}[t!]
  \centering
    \includegraphics[width=0.8\linewidth]{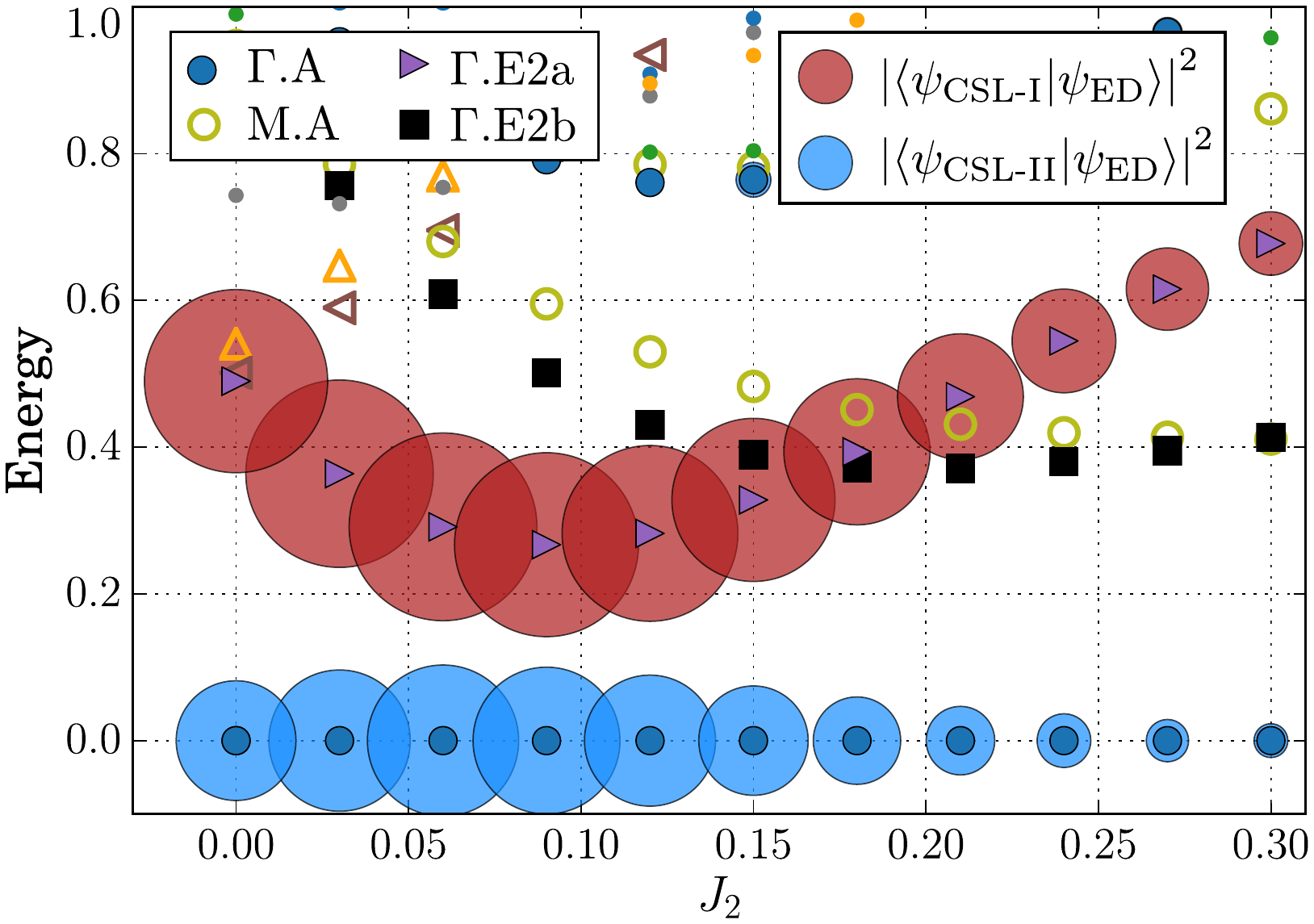}
  \caption{Excitation spectra of the model~\eqref{eq:hamiltonian}
    from ED for $J_\chi = 0.24$ and overlaps with the two CSL
    wavefunctions on a 36 site cluster. Full (empty) symbols denote even (odd)
    spin levels, different types of symbols denote different
    space-group representations. 
    We find overlaps
    $\mathcal{O}_{\textnormal{GW}-\textnormal{ED}}^\alpha$ as in
    Eq.~\eqref{eq:defoverlap} up to $0.92$.
  }
  \label{fig:spectraoverlaps}
\end{figure}

In Fig.~\ref{fig:spectraoverlaps} we show energy spectra for a
horizontal cut in the phase diagram at $J_\chi=0.24$. The first excited
level above the groundstate for $J_2 \lesssim 0.16$ belongs to 
the irreducible representation $\Gamma$.E2b. The region where this 
representation is the first excited state is colored light blue in 
Fig.~\ref{fig:phasediagram}. The parton tight binding model for the 
GPWFs we choose has a two-site unit cell on the triangular lattice with
$\pi/2$ flux through the triangles. This yields a bandstructure with
two bands with Chern numbers $\pm 1$. The groundstate
of this tight binding model at half filling is given by filling 
the orbitals of the lower band. After Gutzwiller projection such 
a state has been shown to yield a CSL wavefunction 
\cite{Mei2015,Wietek2015,Nataf2016,Hu2016}.
To construct the topological partner of the CSL wavefunction
the phases in the tight-binding model before projection can be tuned 
such that locally the flux through each triangle remains $\pi/2$
while the flux through incontractible loops around the torus changes.
The set of fluxes can be chosen arbitrarily, yet after Gutzwiller projection
these states only form a two dimensional space. This can be
verified by computing the overlap matrix for several GPWFs with
different fluxes through the torus. Indeed we find that thereby 
the rank of the overlap matrix is 2 with a numerical precision of 
$\sim10^{-3}$~\cite{Mei2015}. We chose two out of these wave 
functions spanning the CSL subspace and compute the overlaps with
the lowest two numerical eigenstates from ED.
We find that these two model wave functions
$\ket{\psi_{\text{GW}}^\alpha}$ yield very high overlaps 
\begin{equation}
    \mathcal{O}_{\textnormal{GW}-\textnormal{ED}}^\alpha \equiv
    \left|\left<\psi_{\textnormal{ED}}^0|\psi_{\textnormal{GW}}^\alpha\right>\right|^2
    + \left|\left<\psi_{\textnormal{ED}}^1|\psi_{\textnormal{GW}}^\alpha\right>\right|^2
  \label{eq:defoverlap}
\end{equation}
with the two lowest lying eigenstates of ED of up
to $0.92$~
\footnote{Note that both our model wavefunctions do not have a fixed (angular-)
momentum and thus overlap with both exact eigenstates. The fluxes of these two
wavefunctions have been chosen such that one state has mainly overlap
with the first excited state and the other mainly with the
groundstate.}. In Fig.~\ref{fig:spectraoverlaps} we plot the square
overlap $|\braket{\psi_{\textnormal{ED}}^n}{\psi_{\textnormal{CSL}}^\alpha}|^2$
with the respective exact
eigenstate ($n$) as the diameter of the red ($\alpha=1$) and light
blue ($\alpha=2$) circles. The overlaps are large where the first
excited state is in the $\Gamma$.E2b representation and quickly decay 
afterwards. This region coincides approximately with the
region where the CSL model wave function has a low variational energy in 
the upper right panel of Fig.~\ref{fig:orderparameters}. We note that
the CSL phase in this phase diagram is located near a tetrahedral magnetic 
phase, reminiscent of a recent study of a frustrated honeycomb spin 
model~\cite{Hickey2015}. It would be interesting to investigate the nature of the phase 
transitions from the tetrahedral~\cite{Hickey2015} and the $120^\circ$ N\'eel phases into the CSL. Finally 
a recent purely variational study~\cite{Hu2016} also found evidence for a CSL in our 
model for selected values of $J_2$ and $J_\chi$.

\begin{figure}[t!]
  \centering
  \includegraphics[width=0.9\linewidth]{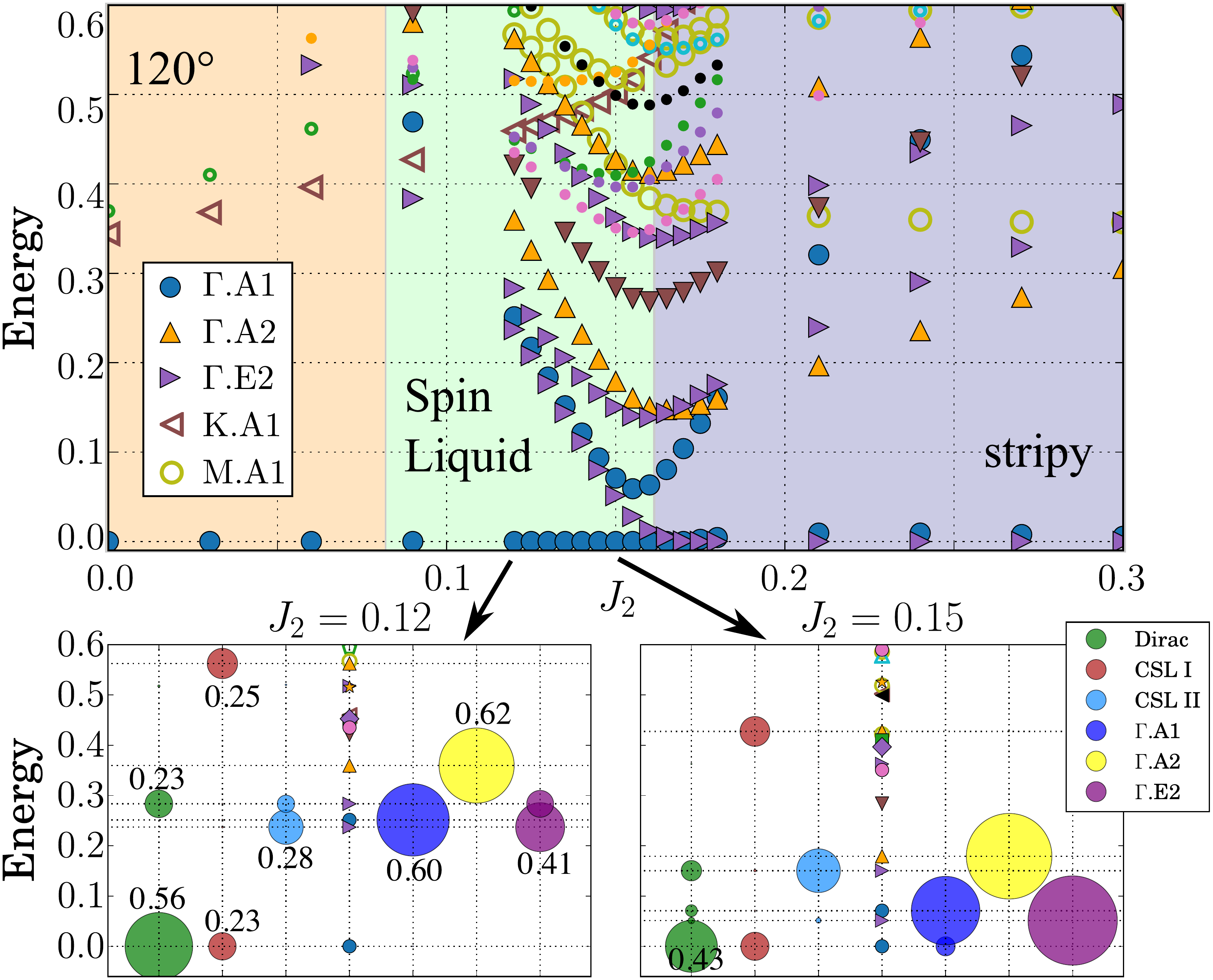}
  \caption{
  ED spectra for $J_\chi = 0$ and spectral decomposition of 
    several model wavefunctions for $J_2=0.12$ and $J_2=0.15$.
    Full (empty) symbols correnspond to even (odd) spin.
    The diameter of the poles  is proportional to the squareoverlap
    $|\braket{\psi_{\textnormal{ED}}}{\psi_{\textnormal{Model}}}|^2$. 
    Besides the CSL and Dirac spin liquid wavefunctions the three 
    wavefuncitons denoted by $\Gamma$.A1, $\Gamma$.A2 and 
    $\Gamma$.E2b are the groundstates in the respective symmetry 
    sectors at $J_2 = 0.3$.  }
  \label{fig:spectraldecomposition}
\end{figure}
\paragraph{Spin disordered state in the $J_1{-}J_2$ Heisenberg model ---} 
We now turn to the time-reversal invariant $J_1{-}J_2$ line with $J_\chi=0$. 
A number of recent numerical works~\cite{Kaneko2014,Zhu2015,Saadatmand2015,Hu2015a,Iqbal2016} involving flavors of variational 
Monte Carlo (VMC)~\cite{Kaneko2014,Iqbal2016} and Density Matrix Renormalization 
Group (DMRG) techniques~\cite{Zhu2015,Saadatmand2015,Hu2015a} found
a spin disordered region between the $120^\circ$ magnetic order region and the
stripy magnetic order at larger $J_2/J_1$. Multiple candidate 
phases for this intermediate parameter range have been
proposed, without a consensus so far. Whereas Ref.~\cite{Zhu2015} proposes a
gapped spin liquid phase, Ref.~\cite{Kaneko2014} proposes an extended gapless ASL
state. In Ref.~\cite{Hu2015a} it was argued that a CSL and a $\mathbb{Z}_2$ spin 
liquid are competing in the low energy sector in the intermediate region 
$0.07 \lesssim J_2 \lesssim 0.15$. Ref~\cite{Iqbal2016} compared
variational energies of several $\mathbb{Z}_2$ spin liquids based on 
Gutzwiller projected wave functions. Interestingly they find that among all
of these wavefunctions the lowest energy is not attained by a state with 
$\mathbb{Z}_2$ structure, but rather by a model whose band structure 
features gapless Dirac-like excitations before projection (see
supp. mat. and Refs.~\cite{Iqbal2016,Lu}). 
After projection 
this state is called \textit{Dirac Spin Liquid} (DSL) and Ref.~\cite{Iqbal2016} finds
an extended gapless region described by a dressed wave function of the DSL kind.

In order to shed light on this open question we present the detailed energy spectrum
of the $N_s=36$ site cluster along the $J_\chi=0$ line in the top panel of Fig.~\ref{fig:spectraldecomposition}.
In the small $J_2$ region the first few levels are in agreement with the tower of state 
expectations for the $120^\circ$ N\'eel state~\cite{Bernu1992}, and similarly at the largest $J_2$ values shown
for the stripy collinear magnetic order~\cite{Lecheminant1995}~\footnote{Some additional levels are visible remnants
of the order by disorder mechanism~\cite{Lecheminant1995}}. 

Focusing on the intermediate region $ 0.08 \lesssim J_2  \lesssim 0.16$ we would expect to
see an approximate four-fold ground state degeneracy in either a non-chiral $\mathbb{Z}_2$
spin liquid or two time-reversal related copies of a CSL as in Refs.~\cite{Gong2014,Wietek2015}.
This is not the case for our system size. 
An additional complication
comes from the observation that some of the low-lying levels in the spin liquid region seem to be states which
become the ground state or low-lying levels in the stripy collinear region across the first order transition around $J_2\sim 0.16$.
 This illustrated by calculating overlaps of several low-lying eigenstates at $J_2=0.3$ with the eigenstates at $J_2=0.12$ ($J_2=0.15$) displayed in 
the lower left (right) panel of Fig.~\ref{fig:spectraldecomposition}.

Given the rather low variational energy of the DSL and to a lesser extent of the CSL model wave functions as shown in the 
right part of Fig.~\ref{fig:orderparameters} (and for the DSL in Refs.~\cite{Kaneko2014,Iqbal2016}) we
also compute the decomposition of these model wave functions onto the exact ED eigenstates for 
$J_2=0.12$ and $J_2=0.15$, as shown in the lower part of Fig.~\ref{fig:spectraldecomposition}. At $J_2=0.12$ 
the ground state has a sizeable overlap with the DSL model wave
function of $0.56$. Furthermore when going up to energies
of about $0.6$, we can also find four states which have non-vanishing overlap with the two different topological sectors
of the CSL model wave functions, although the integrated weight is lower than for the DSL state. This might explain the findings of~Ref.~\cite{Hu2015a}
and is due to the reported CSL stabilized at finite but small $J_\chi$.  In the future one should also 
explore overlaps with a $\mathbb{Z}_2$ spin liquid model wave function in order to address the propensity to this kind of spin liquid on an equal footing
the other model wave functions.

\begin{figure}
  \centering
  \includegraphics[width=\linewidth]{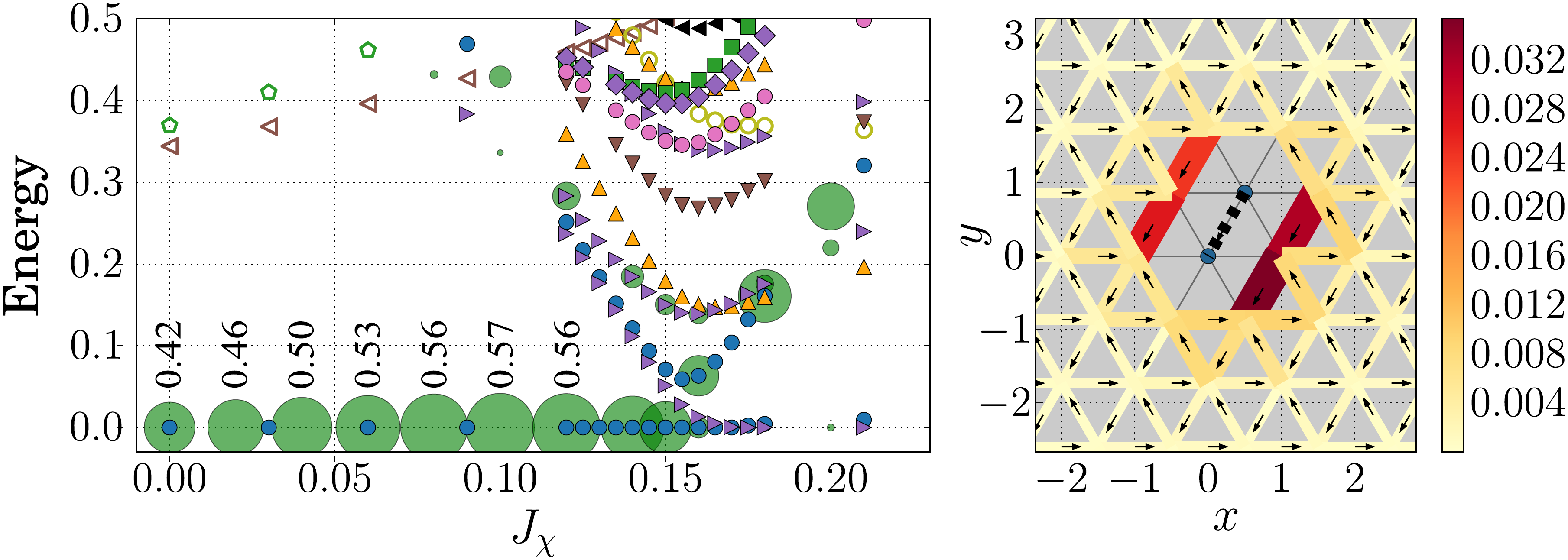}
  \caption{Overlaps of DSL wavefunction with ED eigenstates and decay
    of spin-spin and twist-twist corralation functions $\left< (\vec{S}_0\times \vec{S}_1)\cdot(\vec{S}_i\times \vec{S}_j)\right>$ of the DSL from VMC on a 144 sites lattice.
    The maximum ground state overlap is attained at $J_2=0.1$.
    The correlations decay algebraically over
    distance. 
}
  \label{fig:ovlpcorrdecay}
\end{figure}

We have then explored the overlap of the exact ED ground state with the DSL model wave function in a larger range of $J_2$ couplings and observe the
overlap to be maximal in the vicinity of the putative $120^\circ$
N\'eel to spin liquid quantum phase transition around $J_2\sim 0.08$
in Fig.~\ref{fig:ovlpcorrdecay}.
Motivated by this observation we have explored correlation functions in the DSL model wave function and we find likely power-law correlation functions 
which peak at the $K$ point in reciprocal space (consistent with Refs.~\cite{Kaneko2014,Iqbal2016}). We also investigated the spin vector chirality (twist) 
correlations and find them to exhibit likely power-law correlations with a real space pattern in agreement with the (ordered) pattern in the $120^\circ$ N\'eel 
ordered phase cf. Fig.~\ref{fig:ovlpcorrdecay}.

These nontrivial observations motivate us to conjecture that the DSL wave function should not be considered as a model wave function for an extended ASL region,
but instead as a lattice wave function correctly describing the long-distance properties the quantum critical point out of the $120^\circ$ N\'eel state into a spin liquid. The $O(4)^*$ theory~\cite{Chubukov1994a,Chubukov1994b,Whitsitt2016} is a strong contender describing this transition. Let us put this advocated picture
into a broader context: It is believed that Gutzwiller projected wave functions of partons with $SU(N)$ symmetry and a band structure with $n_D$ Dirac points 
correspond to a lattice realization of QED$_3$: i.e. $N_f=N\times n_D$ two-component Dirac fermions coupled to a compact $U(1)$ gauge field in $2+1$ D. 
It has been shown that in the limit of sufficiently large $N_f$ there are no relevant operators in the theory~\cite{Hermele2004,Hermele2005}, and therefore this 
wave function is representative for an extended ASL region at large $N_f$. For small $N_f<N_f^c$ on the other hand one expects QED$_3$ to become confining in general. The DSL wave function with its power-law decaying correlation functions could then describe a (multi)critical conformal field theory fixed point in between confining phases. The precise value for $N_f^c$ is not known, although recent work~\cite{Grover2014} bounds $N_f^c\lesssim 10$. In the particular case 
of the DSL on the triangular lattice we have $N=2$ and $n_D=2$ resulting in $N_f=4$, substantially lower than the presently known bound. There is also an earlier
observation in Ref.~\cite{Albuquerque2011} that a different $N_f=4$ DSL on the honeycomb lattice describes rather accurately the deconfined quantum critical point~\cite{Senthil1490} between collinear N\'eel order and a VBS phase, giving further evidence that $N_f=4$ DSLs should perhaps be seen as fixed point wave functions for exotic quantum critical points.

The quantum critical scenario naturally comes with divergent correlation lengths, which could be an explanation for the so far missing clear ground state 
degeneracy both in DMRG and ED. Using couplings frustrating both the $120^\circ$ and the stripy N\'eel orders, it might be 
possible to widen the spin-liquid region and to reduce the correlation lengths to numerically accessible scales, allowing to identify the spin liquid unambiguously. It would also be interesting to understand whether the CSL touches the $J_\chi=0$ line at the quantum critical point.

\paragraph{Conclusion ---}
We established the phase diagram of an extended Heisenberg model
on the triangular lattice. Amongst several magnetic orderings 
we found a chiral spin liquid phase in an extended region. 
For the spin disordered region for $J_\chi=0$ we found that the 
DSL has sizeable overlap with ED groundstates. We proposed a scenario
where this wavefunction is the quantum critical wavefunction at 
a transition from magnetic $120^\circ$ N\'{e}el order into a putative
spin liquid phase. 

\acknowledgments
We thank F.~Becca, Y.~Iqbal, S.~Sachdev, M.~Schuler, P.~Strack and S.~Whitsitt for discussions.
We acknowledge support by the Austrian Science Fund 
through DFG-FOR1807 (I-1310-N27) and the SFB FoQus (F-4018-N23).
The computations for this manuscript have been carried out on
VSC3 of the Vienna Scientific Cluster and on the LEOIIIe cluster
of the Focal Point Scientific Computing at the University of Innsbruck.

\bibliography{triangular_j1j2jch}
\clearpage
\appendix

\input{supp_mat_triangular_j1j2jch.tex}

\end{document}

%% file: supp_mat_triangular_j1j2jch.tex
\begin{center}
  \textbf{Supplementary material for Chiral Spin Liquid and Quantum Criticality in extended $S=1/2$ Heisenberg Models on the Triangular Lattice. }
\end{center}

\paragraph{Remarks on Exact Diagonalization calculations:}

\begin{figure}[h]
  \centering
  \includegraphics[width=\linewidth]{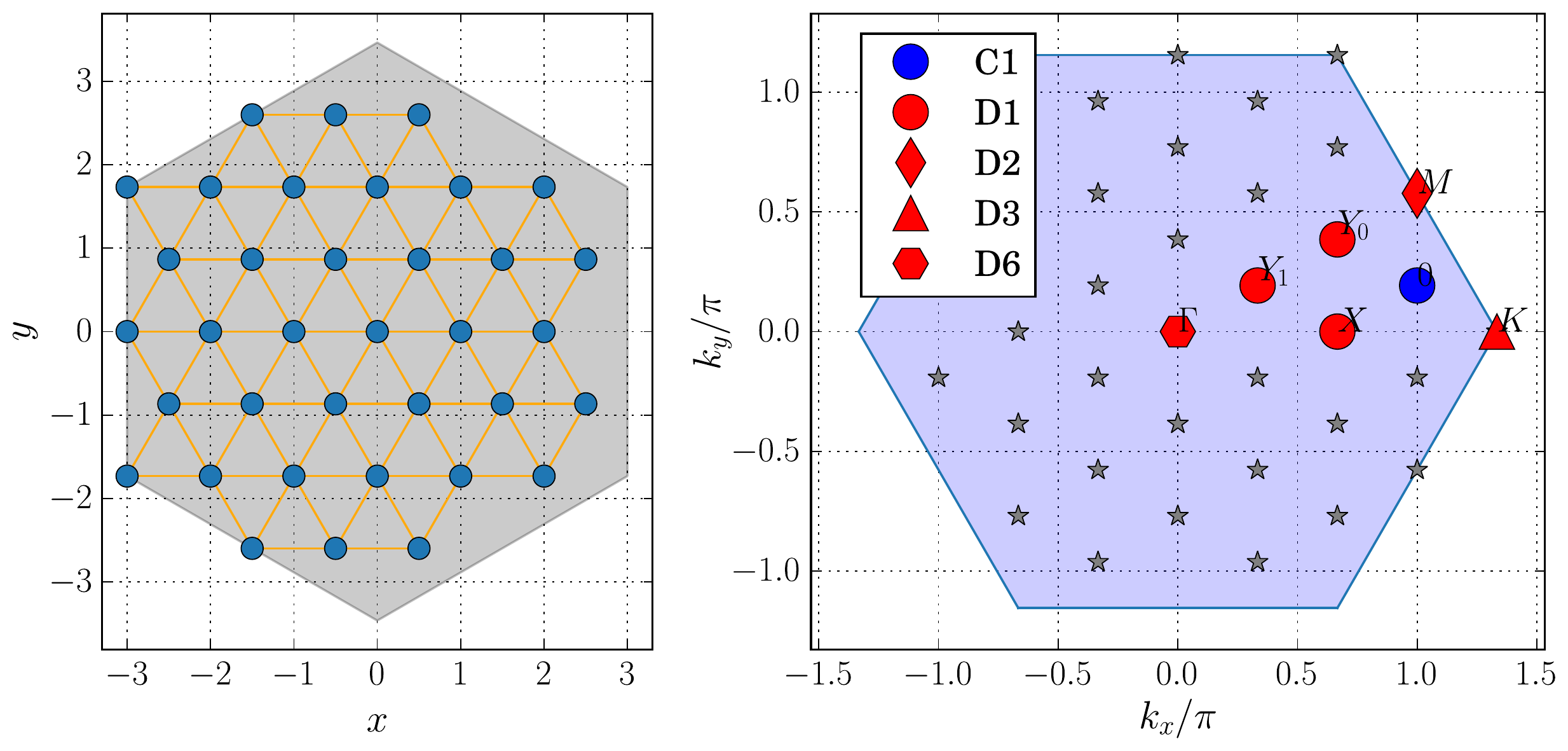}
  \caption{Geometry and Brilluoin zone of simulation cluster with
    $N_s = 36$ used for all simulations in the main text.}
  \label{fig:geometry}
\end{figure}

The simulation cluster we use in our calculations is the $6\times 6$
triangular lattice in Fig.~\ref{fig:geometry} with sixfold rotational
and reflection symmetry. Thus the pointgroup is the full dihedral 
group of order $12$, D6. The momentum space in Fig.~\ref{fig:geometry} 
of this cluster features the $K$ as well as the $M$ point and is thus
capable of stabilizing $120^\circ$, stripy and tetrahedral order. 

The Exact Diagonalization calculations are performed exploiting the
full translational, pointgroup and spinflip symmetry. We thus work in
the basis of generalized Bloch states of the form 
\begin{align*}
  \label{eq:defsymmetrizedstate}
  &\ket{\vec{\sigma}_{\textnormal{sym}}}_{\pm\text{flip},\rho,\vec{k}}=\\
  &\frac{1}{\mathcal{N}}\sum\limits_{s=\pm} \sum\limits_{p\in
    \text{LG}_{\vec{k}}}\sum\limits_{\vec{t}}
  \chi_{\pm\text{flip}}(s)\chi_{\rho}(p)
  \textnormal{e}^{i\vec{k}\cdot\vec{t}}S\circ P \circ \vec{T}\left(\ket{\vec{\sigma}}\right)
\end{align*}
where $\vec{k}$ is the momentum, $s$ denotes the spinflip operation,
$\chi_{\pm\text{flip}}(s)$ the character of even or odd spinflip symmetry
character, $p \in \text{LG}_{\vec{k}}$ a point group element in the little
group of  $\vec{k}$ and $\chi_{\rho}(p)$ the character of a 
representation $\rho$ of the little group. This gives a valid basis 
for all one dimensional representations $\rho$ of the pointgroup.
The two dimensional representations are considered via working
with a corresponding one dimensional representation of a reduced
Little group. For further details see~\cite{Laeuchli2011}.
We use the standard Mulliken notation~\cite{Mulliken1955} 
for labeling the representations of the point group. For the 
dihedral group we have the representations A1, A2, B1, B2, E1, E2.
A1 and A2 are trivial under rotations and A2 is odd under reflections.
B1 and B2 are odd under $180^\circ$ rotations. 
E1 and E2 are two dimensional representations which split
up into the one dimensional E1a, E1b and E2a, E2b of the cyclic C6
subgroup corresponding to $\pm \pi/3$ angular momentum.
The E1a, E2a and E1b, E2b representations are not degenerate without
time-reversal symmetry as for $J_\chi >0$ in the main text
and are then considered seperately.

\paragraph{Anderson Tower of states for magnetic orders}
Magnetic orderings break continuous SU($2$) symmetry of the original 
Heisenberg model. The breaking of continuous SU($2$) symmetry implies
a so called Anderson tower of states~\cite{Lhuillier2005} whose
excitation energies collapse as $1/N_s$ to the groundstate energy in the
thermodynamic limit. They then form the manifold of degenerate 
groundstates in the thermodynamic limit and appear as low
lying excitations on finite cluster size energy spectra. The quantum numbers
of these states can be predicted by group representation theory. 
For the $120^\circ$ and stripy order this has been done in 
Refs.~\cite{Bernu1994, Lecheminant1995}. The method we used
to reproduce their results and calculate the tower of states for the
tetrahedral order is presented in Appendix B of~\cite{Rousochatzakis2008}. 
The irreducible representations of these states in
our notation for small total spin $S$ is given in table~\ref{tab:andersontower}.

\begin{table}[h]
  \centering
  \begin{tabular}{|l|*{3}{c}c|*{3}{c}c|*{4}{c}|}
    \hline
    &\multicolumn{4}{l|}{$120^\circ$ N\'{e}el} 
    &\multicolumn{4}{l|}{stripy order} 
    & \multicolumn{4}{l|}{tetrahedral order}\\
    \hline\hline
      $S$ & $\Gamma$.A1 & $\Gamma$.B1 & K.A1  & \quad 
              & $\Gamma$.A1 & $\Gamma$.E2 & M.A   & \quad
              & $\Gamma$.A & $\Gamma$.E2a & $\Gamma$.E2b & M.A  \\
    \hline
    0   &  1  & 0 & 0 & \quad & 1 & 1 & 0 & \quad & 1&0&0&0\\
    1   &  0  & 1 & 1 & \quad & 0 & 0 & 1 & \quad & 0&1&0&0\\
    2   &  1  & 0 & 2 & \quad & 1 & 1 & 0 & \quad & 0&1&1&1\\
    3   &  1  & 2 & 2 & \quad & 0 & 0 & 1 & \quad & 1&2&0&0\\
    \hline
  \end{tabular}
  \caption{Multiplicities of irreducible representations in the
    Anderson tower of states for the three magnetic orders on the triangular
    lattice defined in the main text. }
  \label{tab:andersontower}
\end{table}


\begin{figure}[h]
  \centering
  \includegraphics[width=\linewidth]{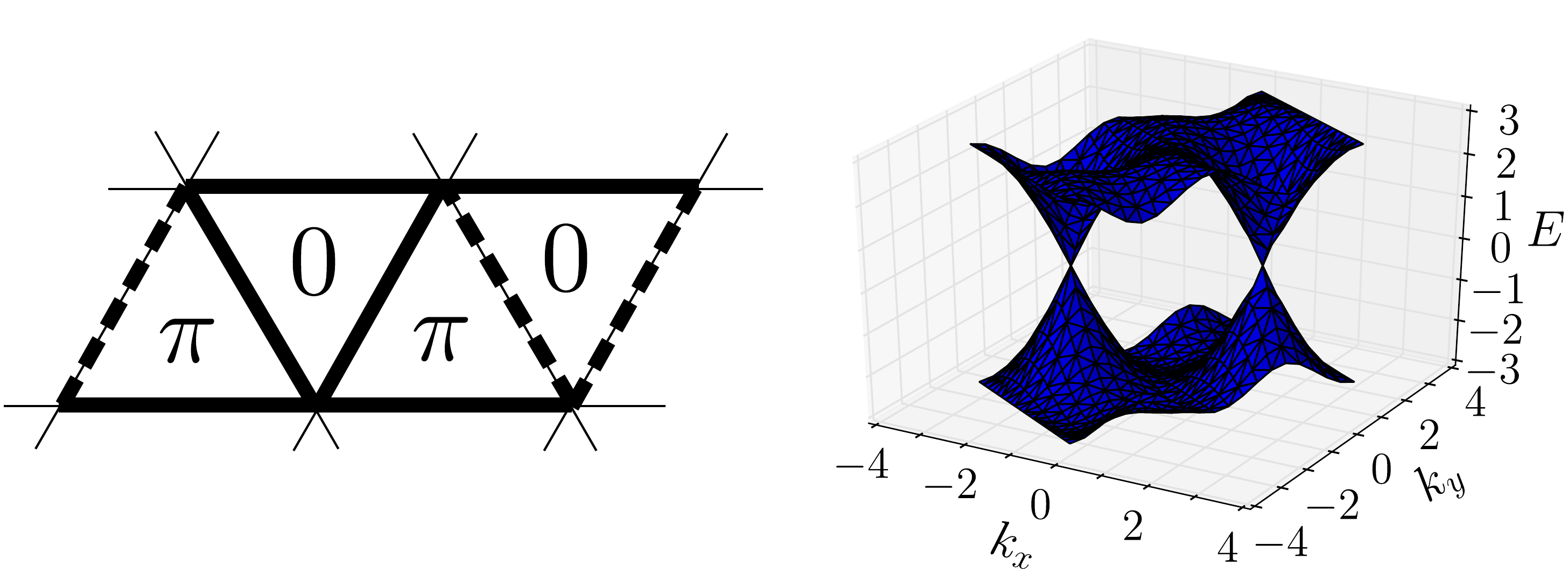}
  \caption{Parton ansatz for the DSL Gutzwiller projected wavefunction
    and parton bandstructure as in Refs.~\cite{Iqbal2016,Lu}. Solid
    (dashed) lines denote real hopping  with amplitude $+1$ ($-1$).}
  \label{fig:decomp_collection}
\end{figure}

\begin{figure}
  \centering
  \includegraphics[width=\linewidth]{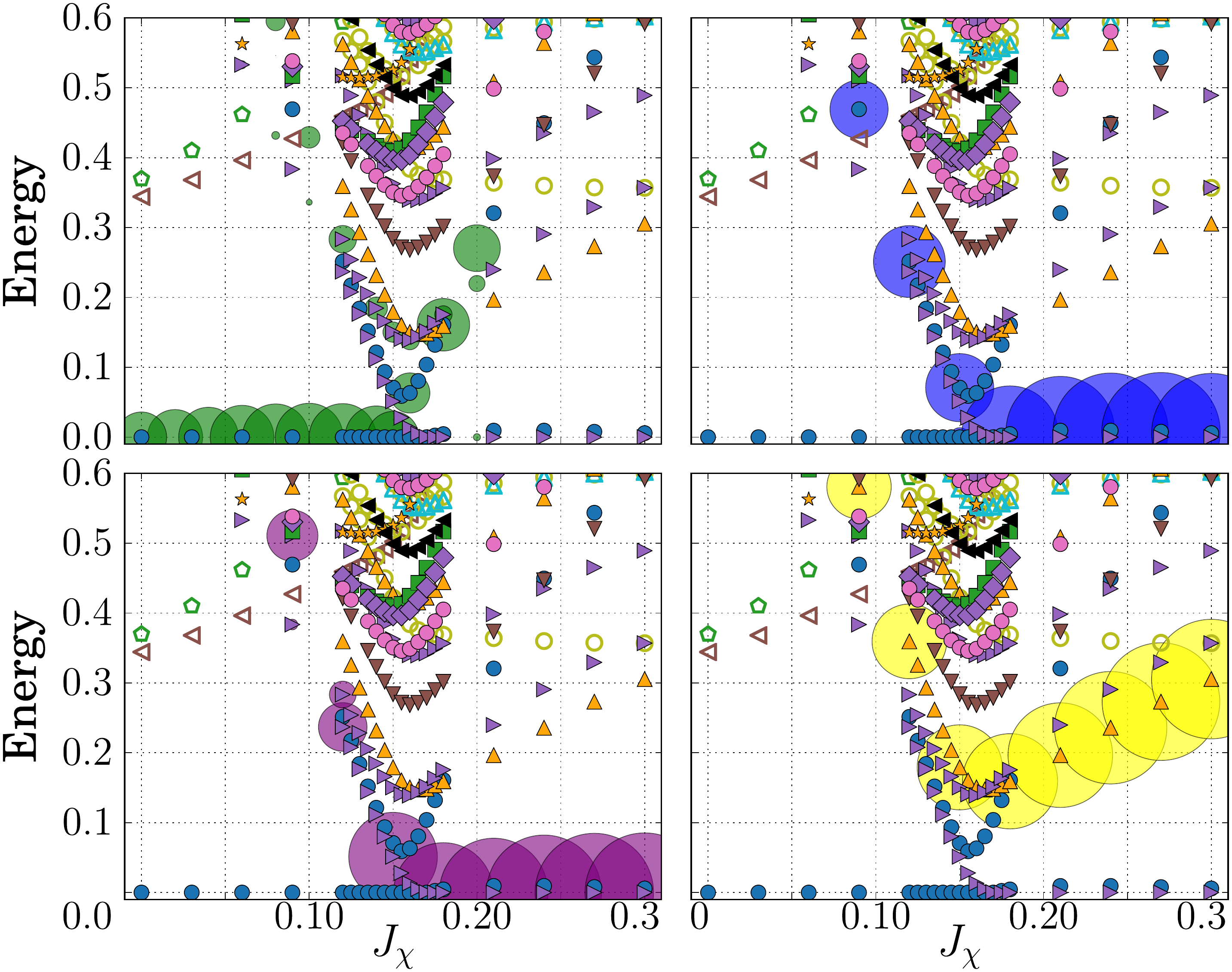}
  \caption{Spectral decomposition of several model wavefunctions on
    the $J_\chi=0$ line. The diameter of the poles  is proportional to the squareoverlap
    $|\braket{\psi_{\textnormal{ED}}}{\psi_{\textnormal{Model}}}|^2$. \textit{Top
    Left:} overlaps with Dirac spin liquid wavefunction. \textit{Top
    right:} overlaps with the groundstate of the $\Gamma$.A1 sector.
  \textit{Bottom Left:} overlaps with the groundstate of the $\Gamma$.E2 sector.
  \textit{Bottom Right:} overlaps with the groundstate of the $\Gamma$.A2 sector.
}
  \label{fig:decomp_collection}
\end{figure}

\begin{figure}[t!]
  \centering
 
  \includegraphics[width=\linewidth]{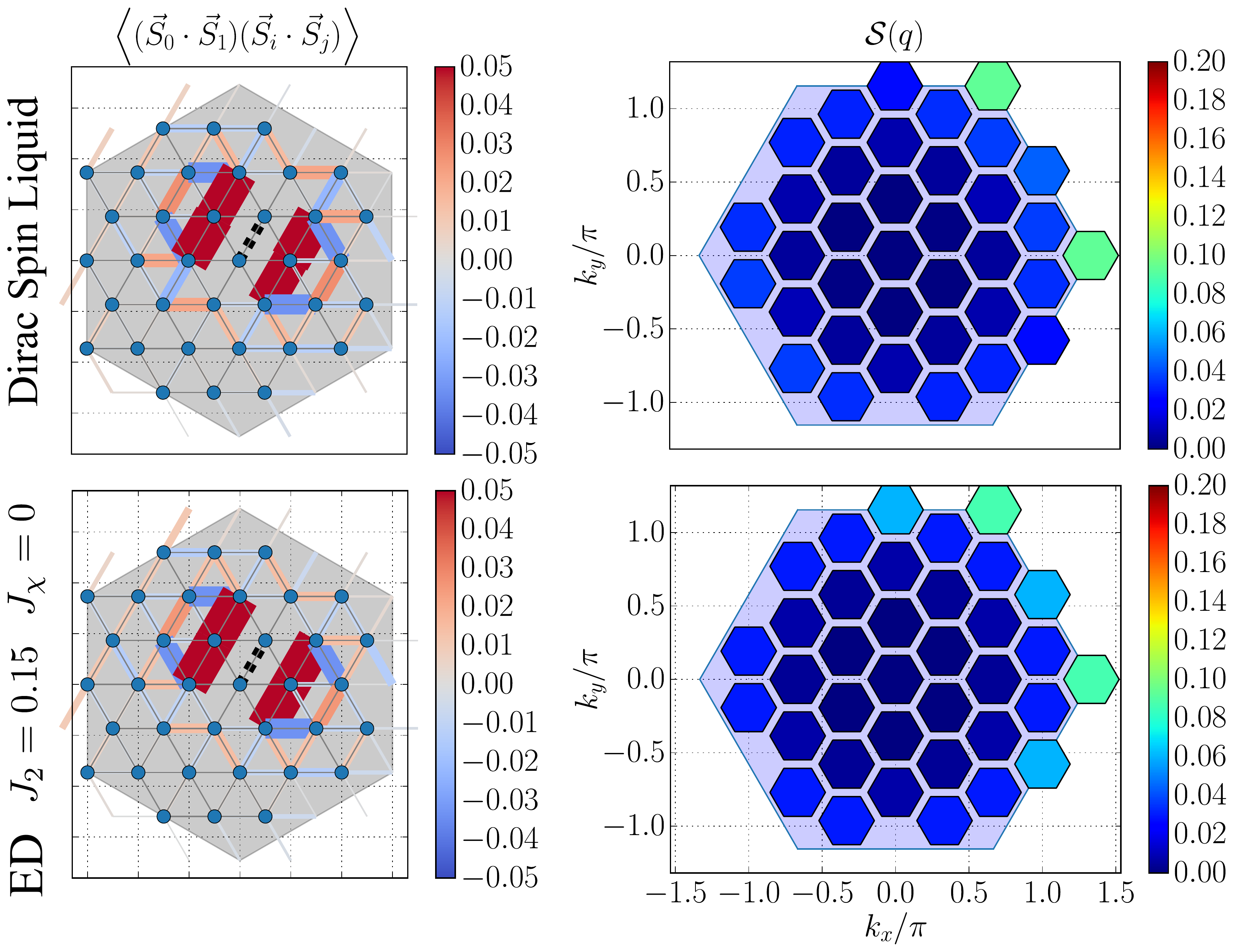}
  \caption{Comparision of connected dimer-dimer correlation functions
    and static spin structure factor between DSL and ED groundstate.
    We see good agreement for the dimer-dimer correlations but
    slight deviations in the spin structure factor due to the onset of
    stripy order in the ED groundstate.}
  \label{fig:correlation_functions}
 \end{figure}